\documentclass[aps,showpacs,preprintnumbers,amsmath, amssymb,nofootinbib]{revtex4}

\usepackage{float}
\usepackage{graphics,epsfig}
\usepackage{bm}
\usepackage{slashed}
\usepackage{graphicx}
\usepackage{amsmath}
\usepackage{amsfonts}
\usepackage{amssymb}
\usepackage{color}%
\usepackage{dcolumn}
\providecommand{\U}[1]{\protect\rule{.1in}{.1in}}

\begin{document}
\baselineskip=0.6 cm
\title{Transport coefficients from hyperscaling violating black brane: shear viscosity and conductivity}
\author{Xiao-Mei Kuang $^{1}$}
\email{xmeikuang@gmail.com}
\author{Jian-Pin Wu $^{2,3}$}
\email{jianpinwu@mail.bnu.edu.cn}
\affiliation{$^1$ Instituto de F\'isica, Pontificia Universidad Cat\'olica de Valpara\'iso, Casilla 4059, Valpara\'iso, Chile
\\$^2$ Institute of Gravitation and Cosmology, Department of Physics, School of Mathematics and Physics, Bohai University, Jinzhou 121013, China
\\ $^3$ State Key Laboratory of Theoretical Physics, Institute of Theoretical Physics, Chinese Academy of Sciences, Beijing 100190, China}
\vspace*{0.2cm}
\begin{abstract}
\baselineskip=0.6 cm
\begin{center}
{\bf Abstract}
\end{center}
We investigate two transport coefficients,  shear viscosity and conductivity, in a non-relativistic boundary filed theory
without hyperscaling symmetry, which is dual to a bulk charged hyperscaling violating black brane. Employing matching method, we
obtain that the ratio of shear viscosity and the entropy density is alway $1/4\pi$ at any temperature,
which satisfies the Kovtun-Starinets-Son (KSS) bound. Besides, we also present the universal formulas of AC conductivity, which is closely dependent on the relation between geometrical parameters $z$ and $\theta$. The optical conductivity at high frequency limit behaves with a (non)-power law scaling or approaches to be constant, depending on the choice of $z$ and $\theta$.
This feature is different from the observes in Lifshitz black brane that  the optical conductivity always complies with a (non)-power law scaling in high frequency limit when the Lifshitz exponent $z>1$.
We also argue that the temperate has no print on the exponent of (non)-power law scaling in large frequency while it will affect the strength of conductivity at low frequency.
\end{abstract}

\maketitle
\newpage
\vspace*{0.2cm}

\section{Introduction}
Gauge/gravity duality is a beautiful approach to study the physics of  strongly coupling sectors,
because it connects a bulk gravitational theory and quantum field theory that lives in one less
dimensions \cite{Maldacena,Gubser,Witten}.  This allows us to explore the strongly coupled
phenomena with the use of dual gravitational systems with weak coupling.
In order to capture physics in a wider
class of field theories, the duality has been generalized to the sectors beyond relativistic conformal symmetry.
A remarkable generalization proposed in \cite{Dong:2012se,Huijse:2011ef,Shaghoulian:2011aa,Alishahiha:2012qu,Papadimitriou:2014lia,Fan:2015aia,Charmousis:2010zz,Chemissany:2014xsa} is to consider the dual gravity with the metric
\begin{eqnarray}\label{HyperscalngGeometry}
ds_{d+1}^2=u^{\frac{2\theta}{d-1}}\left(-\frac{1}{u^{2z}}dt^2+\frac{1}{u^2}du^2+\frac{1}{u^2}d\vec{x}^2\right),
\end{eqnarray}
which presents both a Lifshitz dynamical critical exponent $z$ ($z\geq 1$) and a hyperscaling violating (HV) exponent $\theta$.
Under the scale-transformation $t\rightarrow\lambda^zt,\,\, x_i\rightarrow\lambda x_i,\,\,u\rightarrow\lambda u~$,
the metric transforms as $ds\rightarrow\lambda^{\theta/(d-1)}ds$, which breaks the scale-invariance.
When $\theta=0$, the above analysis recovers the known geometry with Lifshitz symmetry and it goes back to the pure AdS geometry
when $z=1$ and $\theta=0$.
Lots of extensive holographic study based on HV background have been
present in \cite{Lucas:2014sba,Lucas:2014zea,Kuang:2014pna,Kuang:2014yya,Wu:2015schy,Bueno:2014oua,Chen:2015azo,Pan:2015lit,Zhang:2015dyz,Zhou:2015dha,Roychowdhury:2015cta} and references therein.

As the simplest implement of holography, AdS/CFT correspondence has been widespread used in the study
of hydrodynamic properties, such as transport coefficients of strongly coupled systems. Specially,
it is found that the ratio of the shear viscosity ($\eta$) over the entropy density ($s$) has a universal value $1/4\pi$ in dual theories described by Einstein gravity \cite{Policastro:2001yc,Policastro:2002se},
which has been extended into more general theories, see \cite{Maeda:2006by,Cai:2008in,Benincasa:2005qc,Ge:2014eba} and therein.
It is then addressed in \cite{Kovtun:2003wp,Kovtun:2004de} that the value of ratio $1/4\pi$ gives a universal lower bound,
namely the KSS bound, which should be satisfied by all sectors in nature \footnote{However,
we would also like to point out that the viscosity bound will be violated in the presence of higher-derivative gravity corrections \cite{Kats:2007mq,Brigante:2007nu,Brigante:2008gz,Neupane:2009zz,Cai:2008ph,Cai:2009zv,Ge:2008ni,Ge:2009eh,Ge:2009ac,Bhattacharyya:2014wfa,Sadeghi:2015vaa} and in anisotropic gauge/gravity dualities\cite{Rebhan:2011vd,Giataganas:2012zy,Roychowdhury:2015cva}.}.
All the description above are focused on finite temperature.
Later, by borrowing the matching method proposed in \cite{Faulkner:2009wj} to study the holographic (non-)Fermi liquid,
the transport coefficients including shear viscosity and electric conductivity were investigated at extremal AdS RN black hole with finite
charge density \cite{Edalati:2009bi,Edalati:2010hk}.
At zero temperature, the ratio $\eta/s$ of the boundary field theory is the same as that at the finite temperature boundary field theory.
With the same method, the transport coefficients of field theory dual to  AdS charged Gauss-Bonnet is performed in \cite{Cai:2009zn}.

It will be interesting to explore the transport coefficients of hyperdynamic modes in wider boundary geometries like
Eq.(\ref{HyperscalngGeometry})  accompanying with finite charge density which is more general than that discussed in
 AdS gravity.  It implies that one requires  a charged black hole solution with the asymptotical behavior of Eq.(\ref{HyperscalngGeometry})
 in the bulk theory. This kind solution is firstly addressed in \cite{Alishahiha:2012qu}, which will be reviewed in the next section.

Thus, one task of this work is to disclose the shear viscosity of the field theory with finite charge density at any temperatures,
dual to the bulk theory with hyperscaling violating proposed in \cite{Alishahiha:2012qu}.
Before computing the shear viscosity, we calculates the asymptotical solutions of the perturbation modes by matching the solutions at the matching region. The ratio of shear viscosity to entropy density keeps $1/4\pi$ both at zero  and finite temperature, though
the HV exponent $\theta$ explicitly contributes to the entropy density and shear viscosity. This means the hyerscaling violating in a sector does not violate the KSS bound.

The other task is to explore the optical conductivity of this charged HV black brane.
It is addressed in \cite{Hartnoll:2009ns} that the electric conductivity at large frequency in $2+1$ dimensional field theory dual to an AdS geometry approaches to be a constant. Later, the authors of \cite{Sun:2013wpa,Sun:2013zga} argued that  in the four dimensional charged Lifshitz black brane, when the Lifshitz exponent $z>1$, the AC conductivity behaves with a (non-)power scaling in large frequency
limit which is analogous to the phenomena found in some disorder realistic materials \cite{Dyre:2000}. So it is of great interest to further investigate the AC conductivity with $z>1$ and
non-vanishing HV exponent $\theta$. We find that $\theta$ suppress the power law exponent. Especially, even with $z>1$,
certain choice of $\theta$ can make the large frequency behavior of AC conductivity
recover to be a constant. It is because, besides the Lifshitz exponent, the hyperscaling parameter leads to that the UV geometry is different from AdS,  which mainly determine the large frequency behavior of the correlated function of dual operators. Besides, the behavior of (non-)power scaling or constant in
large frequency will not be affected by the temperature because high  frequency always dominate  in energy. However, the lower temperature will suppress the conductivity at low frequency.

This paper is organized as follows. We briefly review the black brane solution at any temperature in HV theory in section \ref{SecGeometry}.
In section \ref{SecEquation}, we present the perturbation equations of the modes. Using the matching method, we study the shear viscosity at zero and non zero
temperature in section \ref{SecViscosity}. In both cases, the KSS bound is satisfied. In section \ref{SecConductivity}, we discuss our numerical results of AC
conductivity. The last section is the conclusion and discussion.

\section{The charged HV black branes from Einstein-Dilaton-Maxwell theory}\label{SecGeometry}
We start from Einstein-Dilaton-Maxwell (EDM) action in $3+1$ spacetime dimensions \cite{Tarrio:2011de}
\begin{equation}\label{EMDaction}
S_g=-\frac{1}{16\pi G}\int \mathrm{d}^{4}x \sqrt{-g}\left[R -\frac{1}{2}(\partial \phi)^2+V(\phi)-
\frac{1}{4}\left(e^{\lambda_1\phi}F^{\mu\nu}F_{\mu\nu}+e^{\lambda_2\phi}\mathcal{F}^{\mu\nu}\mathcal{F}_{\mu\nu}\right)\right].
\end{equation}
The action contains two $U(1)$ gauge fields coupled to a dilaton field $\phi$.
The $U(1)$ field $A$ with field strength $F_{\mu\nu}$ is required to have a charged solution,
while the other gauge field $\mathcal{A}$ with field strength $\mathcal{F}_{\mu\nu}$ and the dilaton field are necessary to generate an anisotropic scaling.
Here $\lambda_1$,$\lambda_2$ are free parameters of this model, which will be determined later.
We can deduce the equations of motion for all the fields from the above action. The Einstein equation of motion for the metric is
\begin{eqnarray}\label{EinsteinEquation}
R_{\mu\nu}-\frac{1}{2}Rg_{\mu\nu}=\frac{1}{2}\partial_{\mu}\phi\partial_{\nu}\phi-\frac{V(\phi)}{2}g_{\mu\nu}
+\frac{1}{2}\left[e^{\lambda_1\phi}(F_{\mu\rho} F_\nu^{\rho} -\frac{g_{\mu\nu}}{4}F^{\rho\sigma}F_{\rho\sigma})+e^{\lambda_2\phi}(\mathcal{F}_{\mu\rho}\mathcal{F}_\nu^{\rho} -\frac{g_{\mu\nu}}{4}\mathcal{F}^{\rho\sigma}\mathcal{F}_{\rho\sigma})\right].
\end{eqnarray}
The equations of motion for the dilaton field is
\begin{eqnarray}\label{KG-DilatonEquation}
\nabla^2\phi=-\frac{dV(\phi)}{d\phi}
+\frac{1}{4}\left(\lambda_1e^{\lambda_1\phi} F^{\mu\nu}F_{\mu\nu}+\lambda_2e^{\lambda_2\phi} \mathcal{F}^{\mu\nu}\mathcal{F}_{\mu\nu}\right).
\end{eqnarray}
The Maxwell equation for the gauge fields are
\begin{eqnarray}
\label{MawellEquation1}
\nabla_\mu\left(\sqrt{-g}e^{\lambda_2\phi}\mathcal{F}^{\mu\nu}\right)&=&0,\\ \label{MawellEquation2}
\nabla_\mu\left(\sqrt{-g}e^{\lambda_1\phi}F^{\mu\nu}\right)&=&0.
\end{eqnarray}
The form of potential $V(\phi)$ plays a very important role in obtaining a charged HV black brane.
Following \cite{Alishahiha:2012qu}, we set $V(\phi)=V_0e^{\gamma\phi}$
with $\gamma$ and $V_0$ being free parameters.
Then  the analytic charged HV black brane solution is \cite{Alishahiha:2012qu}
\begin{eqnarray}\label{BGsolution}
&&
ds_{4}^2= r^{-\theta} \left(-r^{2z}f(r)dt^2+\frac{dr^2}{r^2f(r)}+r^2(dx^2+dy^2)\right),
\\
&&
\label{BGsolutionf}
f=1-\left(\frac{r_h}{r}\right)^{2+z-\theta}+\frac{Q^2}{r^{2(z-\theta+1)}}\left[1-\left(\frac{r_h}{r}\right)^{\theta-z}\right],
\\
&&
\label{BGsolutionFrtmathcal}
\mathcal{F}_{rt}=\sqrt{2(z-1)(2+z-\theta)}e^{\frac{2-\theta/2}{\sqrt{2(2-\theta)(z-1-\theta/2)}}\phi_0}r^{1+z-\theta},
\\
&&
\label{BGsolutionFrt}
F_{rt}=Q\sqrt{2(2-\theta)(z-\theta)}e^{-\sqrt{\frac{z-1+\theta/2}{2(2-\theta)}}\phi_0}r^{-(z-\theta+1)},
\\
&&
\label{BGsolutionephi}
e^{\phi}=e^{\phi_0}r^{\sqrt{2(2-\theta)(z-1-\theta/2)}}.
\end{eqnarray}
Here, $r_h$ is the radius of horizon satisfying $f(r_h)=0$ and $Q=\frac{1}{16\pi G}\int e^{\lambda_1 \phi}F_{rt}$ is the total charge of the black brane.
All the parameters in the action can be determined by Lifshitz scaling exponent $z$ and HV exponent $\theta$
\begin{eqnarray}\label{parameters}
\lambda_1&=&\sqrt{\frac{2(z-1-\theta/2)}{2-\theta}}\nonumber\\
\lambda_2&=&-\frac{2(2-\theta/2)}{\sqrt{2(2-\theta)(z-\theta/2-1)}}\nonumber\\
\gamma&=&\frac{\theta}{\sqrt{2(2-\theta)(z-1-\theta/2)}}\nonumber\\
V_0&=&e^{\frac{-\theta\phi_0}{\sqrt{2(2-\theta)(z-1-\theta/2)}}}(z-\theta+1)(z-\theta+2).
\end{eqnarray}
From Eqs. (\ref{BGsolutionFrtmathcal}) and (\ref{BGsolutionFrt}), one can obtain the gauge fields
\begin{eqnarray}\label{At}
&&
\mathcal{A}_t=-\slashed{\mu}\left[1-\left(\frac{r}{r_h}\right)^{2+z-\theta}\right] ~~\text{with}~~\slashed{\mu}=\frac{\sqrt{2(z-1)(2+z-\theta)}}{2+z-\theta}e^{\frac{2-\theta/2}{\sqrt{2(2-\theta)
(z-1-\theta/2)}}\phi_0}r_h^{2+z-\theta},
\\
&&
\label{RealAt}
A_t=\mu \left[1-\left(\frac{r_h}{r}\right)^{z-\theta}\right]~~\text{with}~~\mu=Q\sqrt{\frac{2(2-\theta)}{z-\theta}}e^{-\sqrt{\frac{z-1+\theta/2}{2(2-\theta)}}\phi_0}r_h^{\theta-z}.
\end{eqnarray}
In addition, the Hawking temperature and entropy density are respectively
\begin{eqnarray}
&&
T=\frac{(2+z-\theta)r_h^z}{4\pi}\left[1-\frac{(z-\theta)Q^2}{2+z-\theta}r_h^{2(\theta-z-1)}\right],
\label{temperature}
\\
&&
s=\frac{r_h^{2-\theta}}{4G}.\label{EntropyDensity}
\end{eqnarray}

Before proceeding, we have to fix the valid  region of
the parameters $z$ and $\theta$. First, the
background solution~(\ref{BGsolution})-(\ref{BGsolutionephi}) are
valid only if $z\geq 1$ and $\theta\geq 0$. Second, the condition $z-\theta\geq
0$ should be satisfied to make sure a well-defined  chemical potential in the dual field theory. Third, it
is obvious from equation~(\ref{At}) that $\theta< 2$.
The null energy condition $(-\frac{\theta}{2}+1)(-\frac{\theta}{2}+z-1)\geq0$ \cite{Alishahiha:2012qu} gives us
$\theta\leq2(z-1)$.
Thus, in this charged background, the range of the parameters is
\begin{eqnarray}\label{ParameterRegion}
\left\{
\begin{array}{rl}
&0\leq \theta \leq 2(z-1) \quad {\rm for} ~~~1\leq z<2   \ ,   \\
&0\leq\theta<2 \quad {\rm for}~~~ z\geq2  \ .
\end{array}\right.
\,
\end{eqnarray}

Furthermore, if we set $Q=\sqrt{\frac{2+z-\theta}{z-\theta}}r_h^{z-\theta+1}$, i.e., $\mu=\frac{\sqrt{2(2-\theta)(2+z-\theta)}}{z-\theta}r_h$,
one reach the zero-temperature limit, in which the redshift factor $f(r)$ becomes
\begin{eqnarray}\label{frTzero}
f(r)|_{T=0}=1-\frac{2(z-\theta+1)}{z-\theta}\left(\frac{r_h}{r}\right)^{z-\theta+2}
+\frac{z-\theta+2}{z-\theta}\left(\frac{r_h}{r}\right)^{2(z-\theta+1)}.
\end{eqnarray}
Obviously, in the $r\rightarrow r_h$ limit,
\begin{eqnarray}\label{frTzerorh}
f(r)|_{T=0,r\rightarrow1}\simeq \frac{(z-\theta+1)(z-\theta+2)}{r_h^2}(r-r_h)^2.
\end{eqnarray}
Therefore, at the zero temperature, there exists the same near horizon geometry, $AdS_2\times \mathbb{R}^2$,
as that of RN-AdS background. Specially, we can define $u=r/r_h$ and change the coordinate via
\begin{equation}\label{u-varsigam}
u-1=\frac{\alpha}{\varsigma},
\end{equation}
with $\alpha=\frac{1}{(z-\theta+1)(z-\theta+2)r_h^z}$, consequently, near the horizon, the metric  can be given by
\begin{eqnarray} \label{MetricNearHorizon}
ds^{2}=r_h^{-\theta}\left[\frac{-dt^{2}+d\varsigma^{2}}{(z-\theta+1)(z-\theta+2)\varsigma^{2}}+r_h^2(dx^{2}+dy^2)\right],
\end{eqnarray}
with the curvature radius of $AdS_2$ $L_2\equiv 1/\sqrt{(z-\theta+1)(z-\theta+2)}$, while the gauge fields are $A_{\tau}=\frac{\mu(z-\theta)\alpha}{\varsigma}$ and $\mathcal{A}_{\tau}=\frac{\slashed{\mu}(2+z-\theta)\alpha}{\varsigma}$, respectively.

\section{The perturbated equations}\label{SecEquation}


We consider the perturbations around the background (Eqs. (\ref{BGsolution})-(\ref{BGsolutionephi}))
\begin{eqnarray}\label{ModesExpandsion}
g_{\mu\nu}=\bar{g}_{\mu\nu}+h_{\mu\nu},~A_\mu=\bar{A}_\mu+a_\mu,~\mathcal{A}_\mu=\bar{\mathcal{A}}_\mu+b_\mu,~\phi=\bar{\phi}+\delta\phi.
\end{eqnarray}
In this paper, we shall only study the shear channel at the limit of $\vec{k}=0$ and so the perturbation $h_{tt}$, $a_t$ and $b_t$ can be turn off.
In addition, we shall work at the radial gauge, in which $h_{u\nu}=0$, $a_u=0$ and $b_u=0$.
Furthermore, we can only turn on the perturbations of $h_{xt}$, $h_{xy}$, $a_x$, and $b_x$
due to the $SO(2)$ symmetry in the $x-y$ plan.
Then, the linearized equations of motions read as
\begin{eqnarray}
\label{eqhxy}0&=&u^2f h^{x''}_{~y}(u)+\left(3 u f + u z f - u \theta f + u^2 f'\right)h^{x'}_{~y}(u)+\frac{\omega^2}{u^{2z}r_h^{2z}f}h^{x}_{~y}(u),\\
\label{eqhxt}0&=&u^4h^{x''}_{~t}(u)+(5 u^3 - u^3 z - u^3 \theta)h^{x'}_{~t}(u)+u^{z+\theta-1} [r_h^{2z-4}(z-\theta)\mu a_x'(u)+r_h^{2\theta-6}(z-\theta+2)\slashed{\mu} b_x'(u)], \\
\label{eqax}0&=&u^2f a^{''}_x(u)+\left(u(3z-\theta-1)f+u^2f'\right)a^{'}_x(u)+
u^{3-3z+\theta}(z-\theta)\mu r_h^{2-2z}h^{x'}_{~t}(u)+\frac{\omega^2}{u^{2z}r_h^{2z}f}a_x(u),\\
\label{eqbx}0&=&u^2f b^{''}_x(u)+\left(u(z+\theta-3)f+u^2f' \right)b^{'}_x(u)+
u^{5-z-\theta}(z-\theta+2)\slashed{\mu}r_h^{2-2z} h^{x'}_{~t}(u)+\frac{\omega^2}{u^{2z}r_h^{2z}f}b_x(u),
\end{eqnarray}
with a constraint equation from the $xu-$ component of linearized Einstein equations
\begin{equation}\label{eqhux}
u^{z+3}h^{x'}_{~t}(u)+u^{2z+\theta-2} [r_h^{2z-4}(z-\theta)\mu a_x(u)+r_h^{2\theta-6}(z-\theta+2)\slashed{\mu} b_x(u)]=0
\end{equation}
where $f=f(u)=1 -[1 +\frac{(z - \theta) \mu^2}{2 (2 - \theta) r_h^2}] (\frac{1}{u}) ^{z -\theta + 2} + \frac{(z -\theta) \mu^2}{2 (2 - \theta) r_h^2}(\frac{1}{u})^{2 (z - \theta + 1)}$ and the prime denotes to the derivative to $u$.
\section{shear viscosity}\label{SecViscosity}
The shear viscosity of the dual boundary theory is related with the retarded Green function via the Kubo formula
\begin{equation}\label{eta}
\eta=-\lim_{\omega\rightarrow0}\frac{\mathbf{Im} G^R_{xy,xy}(\omega)}{\omega}.
\end{equation}
According to the real-time recipe proposed in \cite{Son:2002sd}, the formula of
the retarded Green function is
\begin{equation}\label{recipe}
G^R_{xy,xy}(\omega)=\frac{1}{16\pi G}\sqrt{-g}g^{uu}h^{x*}_{~y}(u)\partial_u h^{x}_{~y}(u)\mid_{u\rightarrow \infty}.
\end{equation}

\subsection{Zero temperature case}
In order to calculate the shear viscosity according to the retarded Green function,
we will firstly work out the asymptotical form of $h^x_{~y}$ at zero temperature by matching the solutions in the inner and outer regions.
For convenience of notation, we will set $h^x_{~y}=\psi$. After introducing
\begin{equation}\label{w-omega}
\mathfrak{w}=\alpha \omega ~~and~~\zeta=\omega\varsigma,
\end{equation}
we have $u=1+\frac{\mathfrak{w}}{\zeta}$ from (\ref{u-varsigam}).
So the matching region near the horizon means taking a double limit $\zeta\rightarrow 0$ and $\mathfrak{w}/\zeta\rightarrow 0$.
Then we rewrite (\ref{eqhxy}) as
\begin{equation}\label{psieq}
u^2f \psi''(u)+\left(3 u f + u z f - u \theta f + u^2 f'\right)\psi'(u)+\frac{((z-\theta+1)(z-\theta+2))^2\mathfrak{w}^2}{u^{2z}f}\psi(u)=0.
\end{equation}
\subsubsection{Solution of inner region}
The perturbation mode near the horizon can be expanded in the low frequency limit as
\begin{equation}
\psi_I(\zeta)=\psi_I^{(0)}(\zeta)+\mathfrak{w}\psi_I^{(1)}(\zeta)+\mathfrak{w}^2\psi_I^{(2)}(\zeta)+\cdots.
\end{equation}
Here the leading term attributes to the near horizon $AdS_2\times \mathcal{R}^2$ geometry. So, in terms of the coordinate $\zeta$, the leading term of (\ref{psieq}) read as
\begin{equation}\label{psiI0-zeta}
\psi_I^{(0)''}(\zeta)+\psi_I^{(0)}(\zeta)=0,
\end{equation}
the general solution of which is
\begin{equation}
\psi_I^{(0)}(\zeta)=a_I^{(0)} \exp({i\zeta})+b_I^{(0)} \exp({-i\zeta}).
\end{equation}
Keep the regularity in mind, we intend to choose the in-going boundary condition which requires to
set $b_I^{(0)}=0$ to cancel the out-going branch.

Then, near the matching region, i.e., in the limit of $\zeta\rightarrow 0$, the in-going result can be expanded as
\begin{eqnarray}
\psi_I^{(0)}(\zeta)\mid_{\zeta\rightarrow 0}\simeq a_I^{(0)}(1+i\zeta)=a_I^{(0)}\left[1+\mathcal{G}_R(\mathfrak{w})\frac{1}{u-1}\right]
\end{eqnarray}
where we have  recalled $\zeta=\frac{\mathfrak{w}}{u-1}$ to express in the coordinate $u$ in the second equlity. Importantly,
\begin{equation}
\mathcal{G}_R(\mathfrak{w})=i\mathfrak{w}
\end{equation}
is just the retarded Green function of a zero-charge scalar operator with conformal dimension one in the IR conformal field theory, of which the dual field is $\psi_I^{(0)}$ in the framework of AdS/CFT correspondence. This is explicit that if one re-scale $\zeta$ into $\zeta/\mathfrak{w}$,  one can see that (\ref{psiI0-zeta}) coincides to the equation of motion for a massless and chargeless scalar field in AdS$_2$ geometry\cite{Faulkner:2009wj}.

Considering the equation (\ref{psieq}) in term of the order of $\mathfrak{w}$, it's easy to find that
the solution in the inner region near the matching region has the form
\begin{equation}\label{Inphi}
\psi_I(u)=a_I^{(0)}+\frac{a_I^{(0)}}{u-1}\mathcal{G}_R(\mathfrak{w})+\cdots
\end{equation}
with the dots denoting the non-vanishing $\mathfrak{w}$ term, which will vanish in the limit of  $\mathfrak{w}\rightarrow 0$.
Note that the solution of inner region is universal, meaning
it does not depend on the geometrical parameters.
We will move on to explore the outer region solution  in order to match the solutions.

\subsubsection{Solution of outer region, matching and the shear viscosity}
Similarly, we expand the perturbation field of the outer region in the low $\mathfrak{w}$ limit as
\begin{equation}
\psi_O(u)=\psi_O^{(0)}(u)+\mathfrak{w}\psi_O^{(1)}(u)+\mathfrak{w}^2\psi_O^{(2)}(u)+\cdots.
\end{equation}
It's straightforward to get the leading order equation of (\ref{psieq})
\begin{equation}\label{Outphieom}
u^2f \psi_O^{(0)''}(u)+\left(3 u f + u  z f - u \theta f + u^2 f'\right)\psi_O^{(0)'}(u)=0,
\end{equation}
from which we see that this equation only explicitly depends on the value of $ \sharp =z-\theta$.

For the general case with $z-\theta=\sharp$,  the general solution of (\ref{Outphieom}) has the form
\begin{equation}\label{Outphi0z2c0}
\psi_O^{(0)}(u)=a_O^{(0)}+\int_1^{r} e^{\int_1^{K[2]}-\frac{2-\sharp-\sharp^2-(2+2\sharp) K[1]^{\sharp}+(\sharp^2+3\sharp) K[1]^{2+2\sharp}}{K[1](2+\sharp-(2+2\sharp) K[1]^{\sharp}+\sharp K[1]^{2+2\sharp})}d K[1]}b_O^{(0)} d K[2],
\end{equation}
where $K[i](i=1,2)$ is the complete elliptic integral of the first kind. Although the above solution is not explicit for general $\sharp$, once
we give the value of $\sharp$, we can obtain an detailed expression of the solution. Furthermore, we can take its  behavior near the matching region and the boundary. Specially, the asymptotical behavior of the above equation has the form
\begin{equation}\label{Outphi0z2c0uInf}
\psi_O^{(0)}(u)|_{u\rightarrow\infty}=(a_O^{(0)}+C_0b_O^{(0)})-\frac{b_O^{(0)}}{(\sharp+2)u^{\sharp+2}}+\cdots
\end{equation}
with the dots denoting the higher subleading terms than $u^{-(\sharp+2)}$, while near the matching region, the behavior of (\ref{Outphi0z2c0}) is
\begin{equation}\label{Outphi0z2c0u1}
\psi_O^{(0)}(u)|_{u\rightarrow1}=-\frac{b_O^{(0)}}{(\sharp+1)(\sharp+2)(u-1)}+a_O^{(0)}+
C_1b_O^{(0)}+\cdots
\end{equation}
where the dots represent the vanishing term in the limit of  $\mathfrak{w}\rightarrow 0$. Note that
$C_0$ and $C_1$ are coefficients only dependent on $\sharp$.
Matching (\ref{Inphi}) and (\ref{Outphi0z2c0u1}) will bring us the following relations
\begin{equation}\label{matchingz2c0}
b_O^{(0)}=-(\sharp+1)(\sharp+2)\mathcal{G}_R(\mathfrak{w})a_I^{(0)},~~~a_O^{(0)}
=\left[1+(\sharp+1)(\sharp+2)C_1)\mathcal{G}_R(\mathfrak{w})\right]a_I^{(0)}.
\end{equation}
Then with the use of the above relations, we can write asymptotic form of $h^{x}_{~y}(u)$ as
\begin{eqnarray}
\nonumber
\psi_O(u)|_{u\rightarrow\infty}&=&a_I^{(0)}\left[1+(\sharp+1)(\sharp+2)(C_1-C_0)\mathcal{G}_R(\mathfrak{w})+\cdots\right]
+(\sharp+1)a_I^{(0)}\mathcal{G}_R(\mathfrak{w})[1+\cdots]u^{-(\sharp+2)}+\cdots\\
&=&a_I^{(0)}\left[1+A(\sharp)\mathcal{G}_R(\mathfrak{w})+\cdots\right]
+(\sharp+1)a_I^{(0)}\mathcal{G}_R(\mathfrak{w})[1+\cdots]u^{-(\sharp+2)}+\cdots
\end{eqnarray}
where in the second line we have defined the constant coefficient $A(\sharp)=(\sharp+1)(\sharp+2)(C_1-C_0)$ . Then taking account of (\ref{eta}) and (\ref{recipe}) and setting $a_I^{(0)}=1$, we obtain the ratio of shear viscosity to the entropy density is
\begin{eqnarray}\label{etasharp}
\nonumber
\eta/s&=&-\lim_{\omega\rightarrow0}\frac{\mathbf{Im} G^R_{xy,xy}(\omega)}{\omega s}\\\nonumber
&=&-\lim_{\omega\rightarrow0}\frac{1}{\omega s} \mathbf{Im}\left[\frac{1}{16\pi G}\sqrt{r_h^{4+2z-4\theta}u^{2+2z-4\theta}}r_h^{\theta}u^{2+\theta}f[-(\sharp+1)(\sharp+2)\mathcal{G}_R(\mathfrak{w})
(1+\mathcal{O}(\mathfrak{w}))u^{-(\sharp+3)}]\right]_{u\rightarrow\infty}\\
&=&\frac{r_h^{2-\theta}}{16\pi G s}=\frac{1}{4\pi}.
\end{eqnarray}
where  we have substituted $\mathcal{G}_R(\mathfrak{w})=i\alpha\omega=\frac{i\omega}{(\sharp+1)(\sharp+2)r_h^z}$ in the second line. Note that the result in AdS case with $z=1$ and $\theta=0$ discussed in \cite{Edalati:2009bi} can be recovered by our case with $\sharp=1$.
So we conclude that in HV background, the the KSS bound $\eta/s=\frac{1}{4\pi}$ always hold at zero temperature,
 independent on the geometrical exponents.
 \subsection{Finite temperature}
At finite temperature, we do not have $AdS_2$ geometry near the horizon because the first oder of the expansion of the redshift
now dominant. But we will take some approximation to work out the
solutions near the horizon as well as the boundary and  match the solutions.
We rewrite equation (\ref{eqhxy}) as
\begin{equation}\label{eqhxyF1}
\psi^{''}(u)+\left(\frac{3+z-\theta}{u}+\frac{f'}{f}\right)\psi^{'}(u)+\frac{\omega^2}{r_h^{2z}u^{2z+2}f^2}\psi(u)=0.
\end{equation}

\subsubsection{Inner region}
In the inner region, we have $u-1\ll 1$ and $f(u)= f'{(r_h)}(u-1)+\cdots$ with $f'(r_h)=4\pi T$, which gives us the leading order of equation as
(\ref{eqhxyF1}) as
\begin{equation}\label{eqhxyF2}
\psi^{''}(u)+\left(\frac{1}{u-1}\right)\psi^{'}(u)+\frac{\omega^2}{(4\pi Tr_h^{z})^2(u-1)^2}\psi(u)=0.
\end{equation}
The solution to the above equation is $\psi(u)=C_1(u-1)^{-i\omega/4\pi T r_h^{z}}+C_2(u-1)^{i\omega/4\pi T r_h^{z}}$. To regularize, we choose the infalling solution by setting $C_2=0$. Then in the low frequency limit,  the solution behaves as
\begin{equation} \label{innerSolution}
\psi(u)=C_1\left(1-\frac{i\omega}{4\pi T r_h^{z}}\log (u-1)\right).
\end{equation}

\subsubsection{Matching region}
The region with $\omega<u\omega\ll 1$ is our matching region, in which the equation becomes
\begin{equation}\label{eqhxyF3}
\psi^{''}(u)+\left(\frac{3+z-\theta}{u}+\frac{f'}{f}\right)\psi^{'}(u)=0.
\end{equation}
Its solution can express as
\begin{equation} \label{matchSolution}
\psi(u)=C_3+C_4\int\frac{du}{f u^{3+z-\theta}}.
\end{equation}
The solution (\ref{matchSolution}) near horizon with $u\rightarrow 1$ and $f(u)= 4\pi T(u-1)$ is
\begin{equation} \label{matchSolutionH}
\psi(u)=C_3+C_4\int\frac{du}{4\pi T(u-1)}=C_3+\frac{C_4}{4\pi T}\log(u-1).
\end{equation}
while at large radius with $u\gg1$ and $f(u)\rightarrow1$, it becomes
\begin{equation} \label{matchSolutionB}
\psi(u)=C_3+C_4\int\frac{du}{u^{3+z-\theta}}=C_3-\frac{C_4}{u^{2+z-\theta}}.
\end{equation}

\subsubsection{Outer region, matching solution and the shear viscosity}
In the asymptotic of the outer region $u\gg1$, it is easy to get $f'(u)\rightarrow0$ and $f(u)\rightarrow1$, so the perturbation equation is
\begin{equation}
\psi^{''}(u)+\left(\frac{3+z-\theta}{u}\right)\psi^{'}(u)+\frac{\omega^2}{r_h^{2z}u^{2z+2}}\psi(u)=0.
\end{equation}
We rewrite the above equation in the coordinate $\mathfrak{u}=1/u$ and obtain the following solution
\begin{equation}
\psi= \left(\frac{\omega \mathfrak{u}^z}{z r_h^z}\right)^{p}\left[C_5 J_{p}\left(\frac{\omega \mathfrak{u}^z}{z r_h^z}\right)
+C_6 Y_{p}\left(\frac{\omega \mathfrak{u}^z}{z r_h^z}\right)\right].
\end{equation}
where $p=\frac{z-\theta+2}{2z}$. According to the feature of Bessel function, in the low frequency limit,  the leading order the above solution is
\begin{equation}\label{outerSolution}
\psi=\tilde{C_6}+\tilde{C_5}\left(\frac{\omega}{r_h^z}\right)^{2p}\mathfrak{u}^{2zp}=\tilde{C_6}+\tilde{C_5}\left(\frac{\omega}{r_h^z}\right)^{\frac{z-\theta+2}{z}}\frac{1}{u^{z-\theta+2}}.
\end{equation}

Now we are ready to match both of inner and outer solutions with those in matching region. This can be achieved by equaling (\ref{matchSolutionH}) with (\ref{innerSolution})  and (\ref{matchSolutionB}) with (\ref{outerSolution}), which gives us the following relations between the coefficients
\begin{eqnarray}
&&\tilde{C_6}=C_3=C_1,~~~~C_4=-\frac{i\omega}{r_h^z}C_1,\nonumber\\
&&\tilde{C_5}\left(\frac{\omega}{r_h^z}\right)^{\frac{z-\theta+2}{z}}=\frac{i\omega}{r_h^z}C_1.
\end{eqnarray}
Thus, the leading order of the asymptotic solution at low frequency limit  is
\begin{eqnarray}
\psi=C_1+C_1\left(\frac{i\omega}{r_h^z}\right)\frac{1}{u^{z-\theta+2}},
\end{eqnarray}
where we will consider the normalizability of the solution near the horizon, i.e. $C_1=1$. Having the above behavior, with the similar algebraic computation using (\ref{eta}) and (\ref{recipe}), it is easy to get the shear viscosity
\begin{eqnarray}
\eta=\frac{r_h^{2-\theta}}{16\pi G}
\end{eqnarray}
which is the same as that in zero temperature, so that the KSS bound is  also fulfilled at finite temperature as we expect.

\section{Conductivity}\label{SecConductivity}
In this section, we turn to the conductivity of the dual boundary theory associated with the real Maxwell field $A$,
so we can turn off the fluctuation of secondary gauge field $\mathcal{A}$.
In addition, due to the symmetry of the two spacial direction, we will only consider the fluctuation of $a_x$ at $\vec{k}=0$, which is dual to a conversed current $J_x$ in the dual boundary theory.
Therefore, combining Eqs.
(\ref{eqax}) and (\ref{eqhux}), we can rewrite the perturbed equation of $a_x$
\begin{equation}\label{axeq}
a_x''(u)+\left[\frac{(3z - \theta-1)}{u}+\frac{ f'(u)}{f(u)}\right]a_x'(u)
+\left[\frac{\omega^2}
{u^{2z+2}f(u)^2}-\frac{2Q^2}{u^{2(z-\theta+2)}f(u)}\right]a_x(u)=0.
\end{equation}
Note that the HV exponent in a $d+2$ dimension geometry
leads to an effective dimension $d_{\theta}=d-\theta$ and in our case, we have $d=2$.
Therefore, we will focus only on the case with $\theta\leqslant 1$ in the following discussion,
so that physics lives in the effective dimension with $d_{\theta}\geqslant 1$ but not  in a fractional space dimension (between $0$ and $1$) \cite{Dong:2012se}.
\subsection{The forms of conductivity in boundary field}
At the boundary,  the asymptotical behavior of $a_x$  depends on the relationship of the $z$ and $\theta$, which is
summarized in Table \ref{table1}.  The coefficients $C_1$ and $C_2$  in the table  can be determined by  numerically integrating
the above Maxwell equation in radial direction.  In the framework of AdS/CFT correspondence,   $C_1$ and $C_2$ are
the source and vacuum expectation value of the operator dual to $a_x$, respectively.
\begin{center}
\begin{table}[h]
\begin{tabular}{|c|c|c|c|}\hline
& $z-\theta<2$   & \begin{minipage}{6cm} \vspace{0.05cm}\centering $C_1+\frac{C_2}{ u^{3z-\theta-2}}$ \vspace{0.05cm}\end{minipage}
& \begin{minipage}{6cm} \vspace{0.05cm}\centering $\sigma=\frac{(3z-\theta-2)C_2}{ i \omega C_1}$ \vspace{0.05cm}\end{minipage}\\
\cline {2 -4}
$\theta<1$   &$z-\theta=2$    &\begin{minipage}{6cm} \vspace{0.05cm}\centering$C_1+\frac{C_1\omega^2 \log u}{2z u^{2z}}+\frac{C_2}{u^{2z}}$  \vspace{0.1cm}\end{minipage}& \begin{minipage}{6cm} \vspace{0.05cm}\centering $\sigma=\frac{(3z-\theta-2)C_2}{ i \omega C_1}
-\frac{\omega}{ i(3z-\theta-2)}$ \vspace{0.05cm}\end{minipage}\\
\cline {2 -4}
&$z-\theta>2$  &\begin{minipage}{6cm} \vspace{0.05cm}\centering$C_1+\frac{C_1\omega^2 }{2z(z-\theta-2)u^{2z}}+\frac{C_2}{u^{3z-\theta-2}}$ \vspace{0.05cm}\end{minipage} & \begin{minipage}{6cm} \vspace{0.05cm}\centering $\sigma=\frac{(3z-\theta-2)C_2}{ i \omega C_1}$ \vspace{0.05cm}\end{minipage}\\
 \hline
& $z-\theta<2$   &\begin{minipage}{6cm} \vspace{0.05cm}\centering$C_1+\frac{C_2} {u^{3z-3}}$  \vspace{0.05cm}\end{minipage} & \begin{minipage}{6cm} \vspace{0.05cm}\centering $\sigma=\frac{(3z-3)C_2}{ i \omega C_1}$ \vspace{0.05cm}\end{minipage}\\
\cline {2 -4}
 $\theta=1$  &$z-\theta=2$     &\begin{minipage}{6cm} \vspace{0.05cm}\centering$C_1+\frac{C_1(\omega^2-2Q^2) \log u}{6u^{6}}+\frac{C_2} {u^{6}}$  \vspace{0.05cm}\end{minipage}& \begin{minipage}{6cm} \vspace{0.05cm}\centering $\sigma=\frac{(3z-3)C_2}{i \omega C_1}
-\frac{\omega^2-2Q^2}{ i(3z-3)\omega}$ \vspace{0.05cm}\end{minipage}\\
 \cline {2 -4}
  &$z-\theta>2$    &\begin{minipage}{6cm} \vspace{0.05cm}\centering$C_1+\frac{C_1(\omega^2 -2Q^2)}{2z(z-3)u^{2z}}+\frac{C_2 }{u^{3z-3}}$  \vspace{0.05cm}\end{minipage} & \begin{minipage}{6cm} \vspace{0.05cm}\centering $\sigma=\frac{(3z-3)C_2}{ i \omega C_1}$ \vspace{0.05cm}\end{minipage}\\ \hline
 \end{tabular}
\caption{\label{table1} The asymptotical behavior of $a_x$ and the expression of conductivity. The coefficients $C_1$ and $C_2$ may depend on $\omega$, which are shown in Table \ref{table2}.}
\end{table}
\end{center}

In order to obtain the conductivity $\sigma_{xx}$  via
\begin{equation}\label{sigma}
\sigma_{xx}=\frac{G^R_{xx}(\omega)}{i \omega},
\end{equation}
we have to fix the quadratic on-shell action for the perturbed Maxwell field, which is
\begin{eqnarray}
S_{a_x}^{(2)}&=&\int dxdydt\left[-\frac{1}{2}u^{z+1}f e^{\lambda_1\phi}a_x a_x'\right]_{u\rightarrow\infty},\nonumber\\
&=&\frac{V_2}{2T}\left[-u^{3z-\theta-1}a_x a_x'\right]_{u\rightarrow\infty},
\end{eqnarray}
where we employ $\int dxdydt=\frac{V_2}{T}$ and $T$ is temperature of the dual field theory.
According to Table \ref{table1}, with different geometry parameters,  $a_x$ behaves differently at  asymptotical boundary,
then  the above expression $S_{a_x}^{(2)}$  has different forms, even with infinite term, so that a suitable counter term may be necessary to
be added into the action to cancel the infinite term \cite{Balasubramanian:1999re,Skenderis:2002wp}.  In the following, we will give the detailed results on the on-shell action  $S_{a_x}^{(2)}$,
the counter term  $S_{ct}$ and  the finite renormalized action  $S_{a_x RN}^{(2)}$ with the related asymptotical behavior of $a_x$.
\begin{itemize}
  \item
  For $ \theta \leqslant 1, z-\theta<2$,
 we have $a_x\sim C_1+\frac{C_2}{u^{3z-\theta-2}}$, which leads to
\begin{eqnarray}
S_{a_x}^{(2)}=\frac{V_2}{2T}\left[(3z-\theta-2)C_1C_2\right].
\end{eqnarray}
In these cases, the above expression $S_{a_x}^{(2)}$ is finite, so we do not need any counter term and $S_{a_x RN}^{(2)}=S_{a_x}^{(2)}$.
  \item
  For $\theta<1, z-\theta=2$, we have $a_x\sim C_1+\frac{C_1\omega^2 \log u}{(3z-\theta-2) u^{3z-\theta-1}}+\frac{C_2}{u^{3z-\theta-2}}$ which leads to
 \begin{eqnarray}
S_{a_x}^{(2)}=\frac{V_2}{2T}\left[(3z-\theta-2)C_1C_2-\frac{C_1^2\omega^2}{3z-\theta-2}+C_1^2\omega^2\log u\right].
\end{eqnarray}
The logarithmic divergence term in $S_{a_x}^{(2)}$ can be cancelled by adding the counter term
 \begin{eqnarray}
S_{ct}^{(2)}&=&-\frac{1}{4}u^{-\theta/2}\log u\int dxdydt \sqrt{-h^{ind}}e^{\lambda_1\phi}(F_{ij}^{ind})^2\nonumber\\
&=&-\frac{V_2}{2T}\left[C_1^2\omega^2\log u\right],
\end{eqnarray}
where $h^{ind}$ and $F_{ij}^{ind}$ are the induced metric and gauge field strength on the boundary, respecitively.
So the finite renormalized action is
 \begin{eqnarray}
S_{a_x RN}^{(2)}=S_{a_x}^{(2)}+S_{ct}=\frac{V_2}{2T}\left[(3z-\theta-2)C_1C_2-\frac{C_1^2\omega^2}{3z-\theta-2}\right].
\end{eqnarray}
\end{itemize}
With the similar calculation, we can obtain the results for the other cases as
\begin{itemize}
 \item
 For $\theta<1, z-\theta>2$:
 \begin{eqnarray}
&&S_{a_x}^{(2)}=\frac{V_2}{2T}\left[(3z-\theta-2)C_1C_2+\frac{C_1^2 \omega^2}{z-\theta-2}u^{z-\theta-2}\right],\nonumber\\
&&S_{ct}^{(2)}=-\frac{1}{4}\frac{u^{-\theta/2}}{z-\theta-2}\int dxdydt \sqrt{-h^{ind}}e^{\lambda_1\phi}(F_{ij}^{ind})^2,\nonumber\\
&&S_{a_x RN}^{(2)}=S_{a_x}^{(2)}+S_{ct}=\frac{V_2}{2T}\left[(3z-\theta-2)C_1C_2\right].
\end{eqnarray}
 \item
 For $\theta=1, z-\theta=2$:
 \begin{eqnarray}
&&S_{a_x}^{(2)}=\frac{V_2}{2T}\left[(3z-3)C_1C_2-\frac{C_1^2(\omega^2-2Q^2)}{3z-3}+C_1^2(\omega^2-2Q^2)\log u\right],\nonumber\\
&&S_{ct}^{(2)}=-\frac{1}{4}u^{-1/2}\log u\left(1-\frac{2Q^2}{\omega^2}\right)\int dxdydt \sqrt{-h^{ind}}e^{\lambda_1\phi}(F_{ij}^{ind})^2,\nonumber\\
&&S_{a_x RN}^{(2)}=S_{a_x}^{(2)}+S_{ct}=\frac{V_2}{2T}\left[(3z-3)C_1C_2-\frac{C_1^2(\omega^2-2Q^2)}{3z-3}\right].
\end{eqnarray}
 \item
 For $\theta=1, z-\theta>2$:
 \begin{eqnarray}
&&S_{a_x}^{(2)}=\frac{V_2}{2T}\left[(3z-3)C_1C_2+\frac{C_1^2 (\omega^2-2Q^2)}{z-3}u^{z-3}\right],\nonumber\\
&&S_{ct}^{(2)}=-\frac{1}{4}u^{-1/2}\left(\frac{1}{z-3}-\frac{2Q^2}{\omega^2(z-3)}\right)\int dxdydt \sqrt{-h^{ind}}e^{\lambda_1\phi}(F_{ij}^{ind})^2,\nonumber\\
&&S_{a_x RN}^{(2)}=S_{a_x}^{(2)}+S_{ct}=\frac{V_2}{2T}\left[(3z-3)C_1C_2\right].
\end{eqnarray}
\end{itemize}

After we have the finite renormalized action with all cases in hands, we can get the Retarded Green function
as $G_{xx}^{R}(x)\sim \frac{2T}{V_2} \frac{\partial(S_{a_x RN}^{(2)})/\partial C_1}{C_1}$. Then recalling equation (\ref{sigma}),
it is easy to deduce the expression of the conductivity $\sigma_{xx}$ with all the cases we discussed above, and
we also list the results in Table \ref{table1}.

\subsection{The numerical results}

Since it is difficult to obtain a general solution of the perturbation equation (\ref{axeq}) in the outer region,
an analytical expression of the conductivity in HV geometry cannot be obtained by standard matching method \cite{Faulkner:2009wj,Edalati:2009bi,Edalati:2010hk}.
So, we will numerically solve it. At the horizon,
we can impose the ingoing condition, i.e., $a_x(u)\sim f(u)^{-i\omega/4\pi T}$. Then we closely depend on the results in Table \ref{table1}
to read off the conductivity.  We try to choose samples of the parameters in all possible cases to
explore the behavior of conductivity.
\begin{figure}[h]
\center{
\includegraphics[scale=0.79]{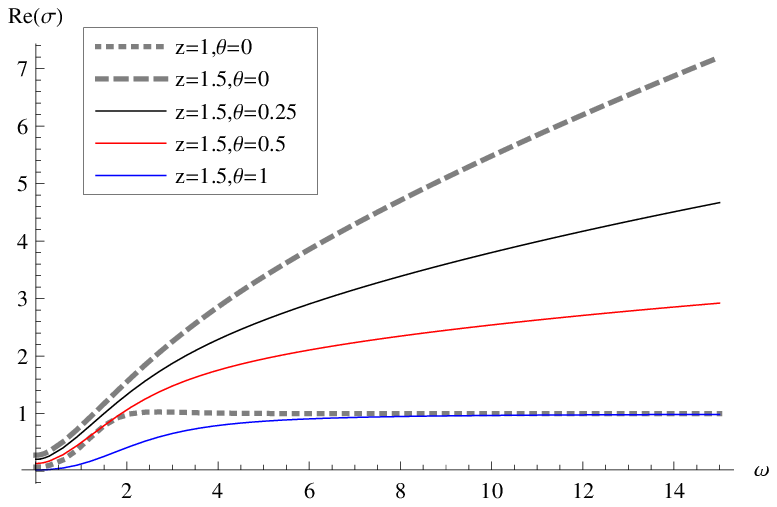}\hspace{0.3cm}
\includegraphics[scale=0.79]{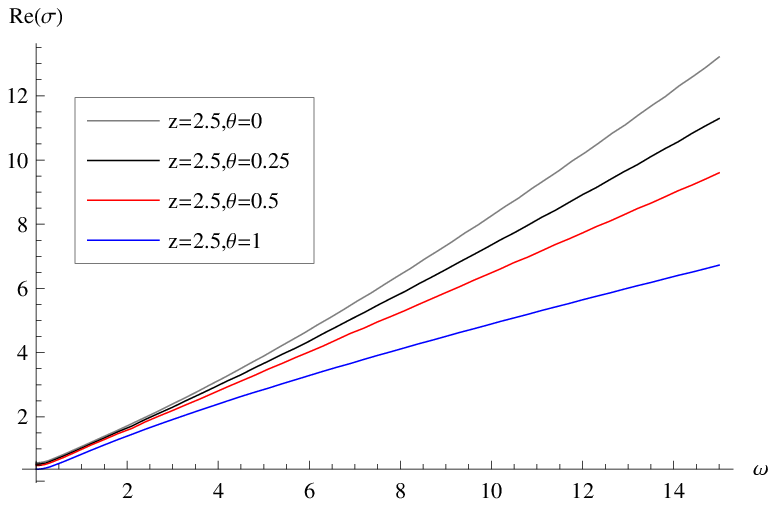}\hspace{0.3cm}
\caption{\label{fig-ReSigma} The real part of conductivity for samples of geometry parameters $z$ and $\theta$ with
fixed temperature $T=1/4\pi$. }}
\end{figure}
\begin{figure}[h]
\center{
\includegraphics[scale=0.8]{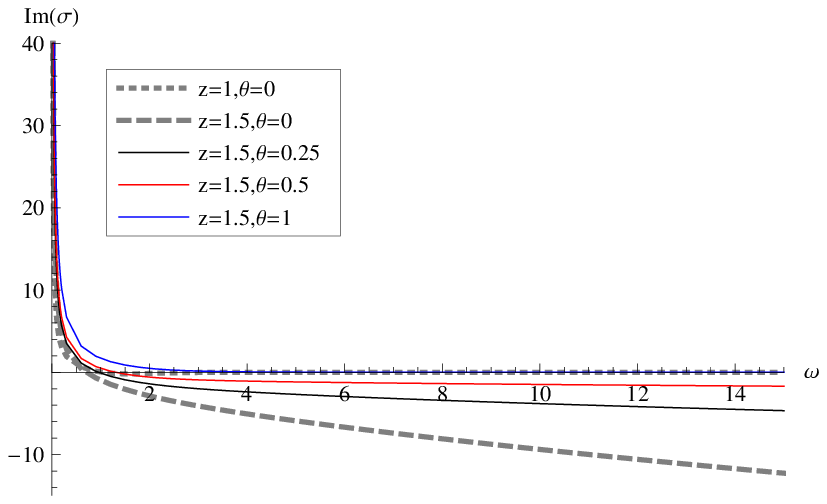}
\includegraphics[scale=0.8]{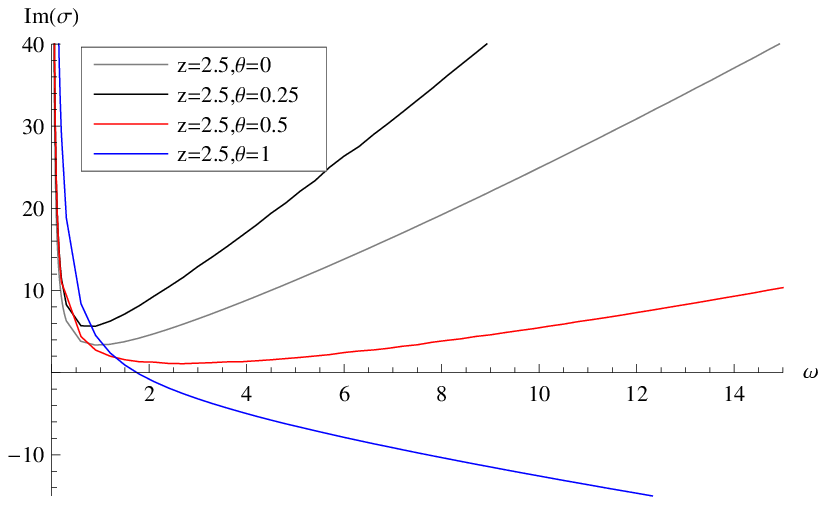}
\caption{\label{fig-ImSigma} The Imaginary part of conductivity for samples of geometry parameters $z$ and $\theta$ with
fixed temperature $T=1/4\pi$.}}
\end{figure}
\begin{center}
\begin{table}
\begin{tabular}{|c|c|c|c|}\hline
z& $\theta$ & \begin{minipage}{6cm} \vspace{0.05cm}\centering Dependence of $C_1, C_2$ in $\omega$ \vspace{0.05cm}\end{minipage} &\begin{minipage}{4cm} \vspace{0.05cm}\centering Large $\omega$ behavior of $\sigma$ \vspace{0.05cm}\end{minipage}\\
\hline
1 & $0$ & \begin{minipage}{6cm} \vspace{0.05cm}\centering $C_1\sim c_1, C_2\sim c_2\omega$ \vspace{0.05cm}\end{minipage} &\begin{minipage}{4cm} \vspace{0.05cm}\centering$\sigma\sim\omega^0$ \vspace{0.05cm}\end{minipage}\\
\hline
&$0$    &\begin{minipage}{6cm} \vspace{0.05cm}\centering  $C_1\sim c_1, C_2\sim c_2 \omega^{5/3}$ \vspace{0.05cm}\end{minipage}&\begin{minipage}{4cm} \vspace{0.05cm}\centering$\sigma\sim\omega^{2/3}$ \vspace{0.05cm}\end{minipage}\\ \cline {2 -4}
& $1/4$     &\begin{minipage}{6cm} \vspace{0.05cm}\centering  $C_1\sim c_1, C_2\sim c_2 \omega^{3/2}$ \vspace{0.05cm}\end{minipage}&\begin{minipage}{4cm} \vspace{0.05cm}\centering$\sigma\sim\omega^{1/2}$ \vspace{0.05cm}\end{minipage}\\ \cline {2 -4}
 $3/2$   & $1/2$     &\begin{minipage}{6cm} \vspace{0.05cm}\centering  $C_1\sim c_1, C_2\sim c_2 \omega^{4/3}$ \vspace{0.05cm}\end{minipage}&\begin{minipage}{4cm} \vspace{0.05cm}\centering$\sigma\sim\omega^{1/3}$ \vspace{0.05cm}\end{minipage}\\ \cline {2 -4}
&$ 1$     &\begin{minipage}{6cm} \vspace{0.05cm}\centering  $C_1\sim c_1, C_2\sim c_2 \omega$ \vspace{0.05cm}\end{minipage}&\begin{minipage}{4cm} \vspace{0.05cm}\centering$\sigma\sim\omega^{0}$ \vspace{0.05cm}\end{minipage}\\
\hline
&$0$    &\begin{minipage}{6cm} \vspace{0.05cm}\centering  $C_1\sim c_1, C_2\sim c_2 \omega^{11/5}$ \vspace{0.05cm}\end{minipage}&\begin{minipage}{4cm} \vspace{0.05cm}\centering$\sigma\sim\omega^{6/5}$ \vspace{0.05cm}\end{minipage}\\ \cline {2 -4}
& $1/4$     &\begin{minipage}{6cm} \vspace{0.05cm}\centering  $C_1\sim c_1, C_2\sim c_2 \omega^{21/10}$ \vspace{0.05cm}\end{minipage}&\begin{minipage}{4cm} \vspace{0.05cm}\centering$\sigma\sim\omega^{11/10}$ \vspace{0.05cm}\end{minipage}\\ \cline {2 -4}
 $5/2$   & $1/2$     &\begin{minipage}{6cm} \vspace{0.05cm}\centering  $C_1\sim c_1, C_2\sim (c_2+c_3\log\omega) \omega^{2}$ \vspace{0.05cm}\end{minipage}&\begin{minipage}{4cm} \vspace{0.05cm}\centering$\sigma\sim(d_1+d_2\log\omega)\omega$ \vspace{0.05cm}\end{minipage}\\ \cline {2 -4}
&$ 1$     &\begin{minipage}{6cm} \vspace{0.05cm}\centering  $C_1\sim c_1, C_2\sim c_2(\omega^2-2Q^2)^{9/10} $ \vspace{0.05cm}\end{minipage}&\begin{minipage}{4cm} \vspace{0.05cm}\centering$\sigma\sim\omega^{4/5}$ \vspace{0.05cm}\end{minipage}\\
 \hline
 \end{tabular}
\caption{\label{table2} Dependence of the coefficient $C_1$ and $C_2$  in Table \ref{table1} on $\omega$ and the large frequency behavior of conductivity.
Here  $c_i (i=1,2,3)$ and $d_i (i=1,2)$ are constants.}
\end{table}
\end{center}
\begin{figure}
\center{
\includegraphics[scale=0.79]{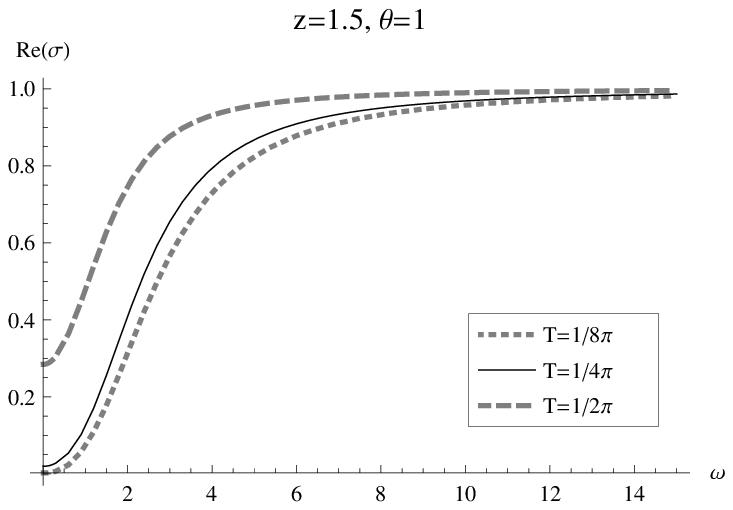}\hspace{0.3cm}
\includegraphics[scale=0.79]{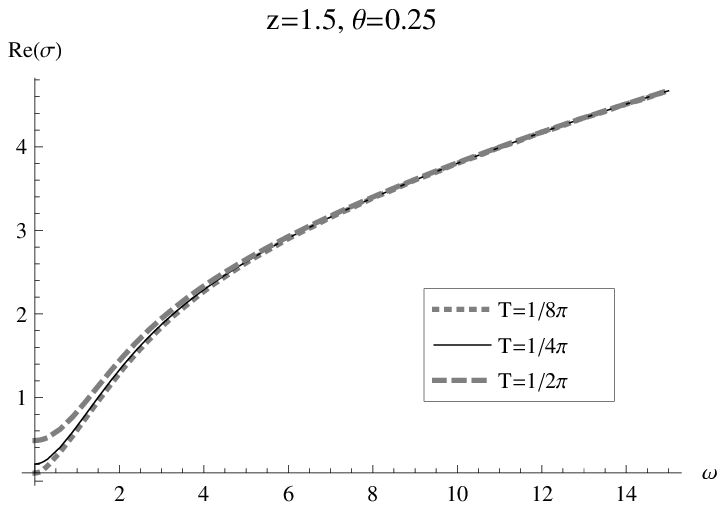}
\caption{\label{fig-T-ReSigma} The real part of conductivity at different  temperatures.}}
\end{figure}
\begin{figure}
\center{
\includegraphics[scale=0.79]{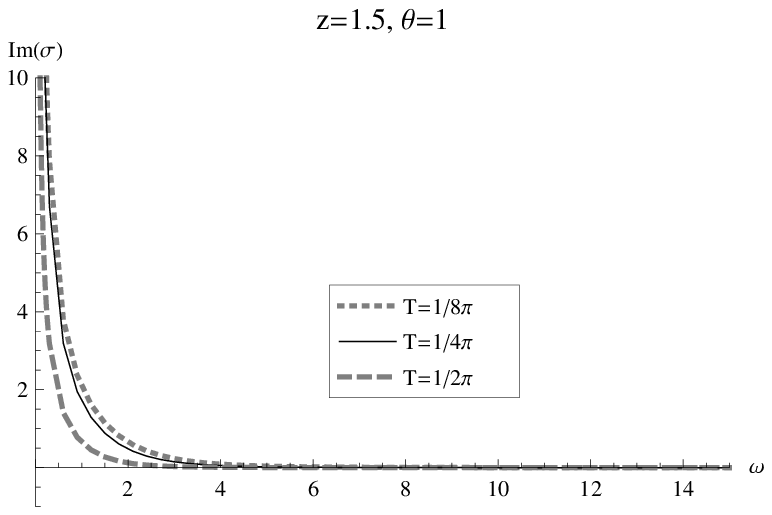}\hspace{0.3cm}
\includegraphics[scale=0.79]{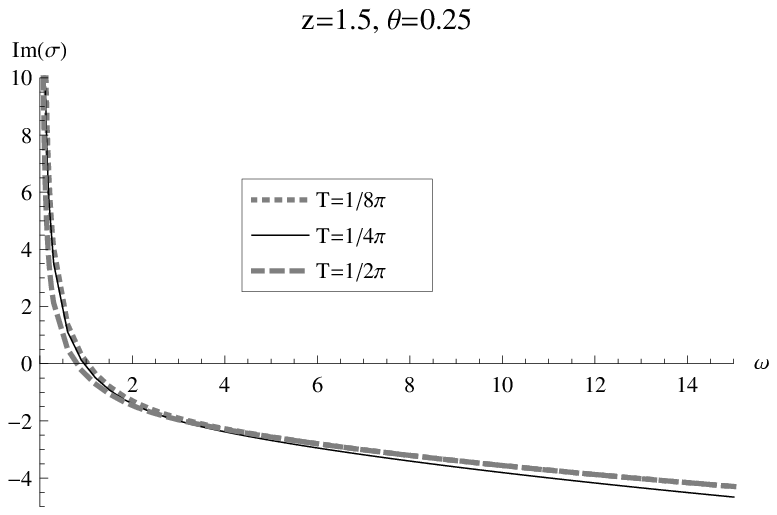}
\caption{\label{fig-T-ImSigma} The imaginary part of conductivity at different  temperatures.}}
\end{figure}
The properties of the conductivity in RN-AdS and charged Lifshitz black hole have been studied in \cite{Edalati:2009bi} and \cite{Sun:2013wpa,Sun:2013zga}, respectively. Here, we shall explore the properties of the conductivity in charged HV black hole.
\begin{itemize}
\item \textbf{Low frequency behavior} FIG.\ref{fig-ReSigma} and FIG.\ref{fig-ImSigma} exhibit the real and imaginary part of conductivity in charged HV black hole.
Their behaviors at low frequency limit follow the same features as that in RN-AdS black hole ($z=1, \theta=0$),
which can be attributed to the same IR geometry shared by RN-AdS and charged HV black hole.
Note that the divergence of the imaginary part at $\omega=0$ indicates a delta function at $\omega=0$
in the real part emerges according to the Kramers-Kronig relations \cite{Hartnoll:2009sz}, which can not be detected by taking the limit of $\omega\rightarrow0$.
\item \textbf{High frequency behavior} 
In \cite{Sun:2013wpa,Sun:2013zga}, it has been shown that the conductivity at high frequency region in non-relativistic CFT dual to Lifshitz gravity follows a power law behavior, i.e., $\sigma\sim\omega^{\delta}$.
However, here in our case, with fixed $z$, HV exponent $\theta$ always suppresses the power law exponent $\delta$  (see FIG.\ref{fig-ReSigma} and FIG.\ref{fig-ImSigma}).
Specially, with fixed $z=3/2$, a constant conductivity, as it occurs in AdS geometry, is recovered when $\theta=1$.
Such peculiar properties of conductivity in high frequency region can be attributes to that the dual boundary theory is non-relativistic field theory with Lifshitz and HV exponents but not a conformal field theory.
Furthermore, we would like to provide an analytical analysis for the high frequency behavior of conductivity.
In Table \ref{table2}, we show the analytical results of the dependent relation of the coefficients $C_1$ and $C_2$ on $\omega$ at large frequency for samples of $z$ and $\theta$, which can give the corresponding behavior of conductivity in terms of Table \ref{table1}.
The analytical results of conductivity at high frequency region are summarized in Table \ref{table2}, which is in perfect agreement with what we observed in FIG. \ref{fig-ReSigma} and FIG. \ref{fig-ImSigma}.
\item \textbf{Temperature dependence} FIG.\ref{fig-T-ReSigma} and FIG.\ref{fig-T-ImSigma} shows the conductivity at different temperatures. It is obvious that with different temperatures,
the behavior of conductivity at the high frequency is almost with the same slope. This is because for $\theta\leqslant 1$, large $\omega$ term of Eq. (\ref{axeq}) near the boundary always dominates, so the change of
temperature, or charge or chemical potential hardly has contribution. It is a universal feature of conductivity in normal and hairy black hole.
While lower temperature will suppress the conductivity at very low frequency region as we expected because lower $T$ lead to
electron inactive so that to weaken the DC conductivity. Though our numeric can not completely touch zero temperature, it can not stop us to deduce that $Re\sigma(\omega\rightarrow0)\sim\omega^2$ and $Im\sigma(\omega\rightarrow0)\sim 1/\omega$ because this low frequency behavior at zero temperature is determined by the IR geometry as revealed in \cite{Edalati:2010hk}.
\end{itemize}

\section{Conclusion and Discussion}
In this paper, by matching method, we have calculated the shear viscosity of a holographic non-relativistic effective field theory at both zero and finite temperature, which is dual to a HV gravity.
We find that the ratio of shear viscosity and the entropy density is always $1/4\pi$ at any temperature as that found in RN-AdS black hole, which satisfies the KSS bound.
Our result shows that the domain of universality of the ratio can be enlarged to include holographic HV effective field theory.

Also, we study numerically the properties of optical conductivity.
Since RN-AdS and charged HV black brane share the same near horizon geometry,
the low frequency behaviors follow the same features.
While the high frequency behaviors of optical conductivity exhibit a (non)-power
law scaling or approaches to be constant, depending on the geometrical parameters $z$ and $\theta$.
It is different from that observed in Lifshitz black brane that the optical conductivity follows a (non)-power law scaling in high frequency limit for $z>1$.

Here, we only calculate the shear viscosity and conductivity from HV black brane at $\vec{k}=0$.
It is worthwhile to calculate the shear and sound modes at finite $\vec{k}$.
We can also calculate the transport coefficients of the other non-relativistic backgrounds, for example, the background with the Schr\"odinger symmetry \cite{Adams:2009dm}.

\begin{acknowledgments}
We are grateful to Wei-Jia Li and Zhenhua Zhou for valuable discussions.
X. M. Kuang is funded by FONDECYT grant No.3150006
and the PUCV-DI Projects No.123.736/2015.
J. P. Wu is supported by the Natural Science Foundation of China under Grant
Nos. 11305018 and 11275208 and also supported by Program for Liaoning Excellent Talents in University (No. LJQ2014123).
\end{acknowledgments}

\end{document}